\documentstyle[12pt]{article}

\begin{document}

\begin{flushright}
IMSc/2004/05/22 \\
hep-th/0405084 
\end{flushright} 

\vspace{2mm}

\vspace{2ex}

\begin{center}
{\large \bf A Description of multi Charged Black Holes \\

\vspace{2ex}

in terms of Branes and Antibranes } \\

\vspace{8ex}

{\large S. Kalyana Rama and Sanjay Siwach \footnote{Name changed
from Sanjay to Sanjay Siwach. Previous papers were written under
the name Sanjay.}}

\vspace{3ex}

Institute of Mathematical Sciences, C. I. T. Campus, 

Taramani, CHENNAI 600 113, India. 

\vspace{1ex}

email: krama@imsc.res.in, sanjay@imsc.res.in \\ 

\end{center}

\vspace{6ex}

\centerline{ABSTRACT} 

\begin{quote} 
We describe multicharged black holes in terms of branes and
antibranes together with multiple copies of gas of massless
excitations. Assuming that energies of these copies of gas are
all equal, we find that the entropy of the brane antibrane
configuration agrees with that of the multicharged black hole in
supergravity approximation, upto a factor $X$. We find that $X =
1$ for a suitable normalisation which admits a simple empirical
interpretation that the available gas energy is all taken by one
single gas which is, in a sense, a certain superposition of the
multiple copies; and that the brane tensions are decreased by a
factor of 4. This interpretation renders superfluous the
assumption of equal energies, which is unnatural from a physical
point of view.
\end{quote}

\vspace{2ex}

\newpage

\vspace{4ex}

{\bf 1.}  
In a recent paper \cite{dgk} Danielsson, Guijosa, and Kruczenski
(DGK) have given, among other things, a description of certain
charged black holes in terms of brane antibrane configurations
which is valid in the far extremal and Schwarzschild regime
also. This has been generalised to other single brane
configurations with or without rotation \cite{g,peet,bl} and to
intersecting multi brane configurations \cite{k}.

In this letter, we study the description of multicharged black
holes in terms of intersecting brane antibrane configurations
\cite{kt',ct,kt,ib}. Following DGK, we obtain the corresponding
multicharged black holes as stacks of intersecting branes and
antibranes, together with massless excitations. Such a stack of
branes and antibranes can be put together (or ``taken apart'')
in $2^{K - 1}$ different ways, along with two copies of gas of
massless excitations for each possibility. ($K$ is the number of
charges.) Therefore, it seems necessary to consider stacks of
branes and antibranes together with $2^K$ copies of gas of
massless excitations not interacting with each other. The
dynamics of such a gas can be obtained as in \cite{dgk} from the
near extremal limit of the corresponding supergravity solutions
\cite{bhreview}.

As we will see in this letter, such a configuration suffices to
describe the dynamics and to obtain the resulting entropy of the
multicharged black hole in the far extremal regime also.
However, in contrast to normal physical situations where one
naturally takes temperatures to be equal, here it is necessary
to assume that the energies of the copies of the gases are
equal. The reason for this assumption, nor a physical mechanism
that can enforce it, is not clear. Nevertheless, we assume this
to be the case and proceed with the analysis as in \cite{dgk}.
We find that the entropy of the brane antibrane configuration
agrees with that of the multicharged black hole in supergravity
approximation, upto a deficit factor $X$.

We analyse the deficit factor $X$ by studying how the resulting
entropy changes if one normalises the brane tension, gas energy,
and the entropy by constant factors. We find that one can indeed
have $X = 1$ for a suitable normalisation which admits a simple
empirical interpretation as in \cite{k}. It implies that the
available gas energy is all taken by one single gas which is, in
a sense, a certain superposition of the $2^K$ copies of the gas
on the brane antibrane stacks; and that the brane tensions are
all to be decreased by a factor of 4. However, the precise
nature of this superposition is not clear to us. If this
interpretation is correct then the entropy of the brane
antibrane configuration is exactly equal to that of the
multicharged black hole in supergravity approximation; and the
assumption that energies, not temperatures, of the different
copies of gas are all equal - an assumption which is unnatural
from a physical point of view - is now rendered superfluous.
 
This letter is organised as follows. In section {\bf 2} we give
a brief description of the relevent results of \cite{dgk}. In
section {\bf 3}, we describe the multicharged black hole in
terms of brane antibrane configurations. In section {\bf 4}, we
give an interpretation of the deficit factor. In section {\bf
5}, we conclude by mentioning a few issues for further study.

{\bf 2.}  
We give a brief description of the relevent results of
\cite{dgk} for the case of single charged black holes. Consider
stacks of $p-$branes which, in supergravity approximation,
correspond to single charged black holes in transverse 
$(d + 3)-$dimensional spacetime. The corresponding 
$(d + 3)-$dimensional charged black holes are obtained as a
system consisting of (i) a stack of branes $N$ in number; (ii) a
stack of antibranes $\bar{N}$ in number; and (iii) a gas of
massless excitations on each stack of branes, with the gas on
different stacks assumed not to interact with each
other. Following DGK, we assume that such a system alone
suffices to describe the dynamics of charged black holes.

The branes and antibranes have zero entropy and energies given
by $C N$ and $C \bar{N}$ respectively where the constant $C$
includes tension and volume of the branes. The gas on the branes
and antibranes have energies $E$ and $\bar{E}$ respectively, and
their entropies $S$ and $\bar{S}$ are given by
\begin{equation}
S = A N^\gamma E^\lambda \; , \; \; \; 
\bar{S} = A \bar{N}^\gamma \bar{E}^\lambda  \; , 
\end{equation}
where $A$ includes brane tension and volume, and $\gamma$ and
$\lambda$ are constants. The net charge $q$, the total energy
$M$, and the total entropy $S_{tot}$ of the system are then
given by
\begin{equation}\label{stot}
q = N - \bar{N} \; , \; \; \; 
M = C (N + \bar{N}) + E  + \bar{E} \; , \; \; \;
S_{tot} = S(E) + \bar{S}(\bar{E}) \; . 
\end{equation}

In canonical formalism, such a system is unstable, for
sufficiently small values of $q$, towards creating an infinite
number of brane antibrane pairs. Hence, one must work in
microcanonical formalism where the net charge $q$ and the total
energy $M$ of the system are kept fixed. In normal physical
systems, one assumes that the total system has one definite
temperature.  But, it turns out that one must instead assume
that $E = \bar{E}$. However, the physical mechanism which
enforces this equality is not well understood. The equilibrium
quantities are then determined by maximising the entropy
$S_{tot}$ of the system with respect to $N$, keeping $q$ and $M$
fixed and setting $E = \bar{E}$.

In supergravity approximation, a stack of $p-$branes describes
well the extremal and near extremal limits of single charged
black holes which have zero entropy in the extremal limit. The
entropy $S(E)$ of the gas on the stack of branes is then the
same as that of the charged black holes in near extremal limit.

Following \cite{kt'}, see also \cite{ib}, consider $p-$branes in
$D = d + p + 3$ dimensional spacetime, which will correspond to
charged black holes in $D - p = d + 3$ dimensional spacetime.
Let $\gamma$ and $\lambda$ be given by
\begin{equation}\label{lambda}
\gamma = \frac{2 (D - 2)}{2 d (p + 1) + a^2 (D - 2)}
\; , \; \; \;
\lambda = \frac{d + 1}{d} - \gamma
\end{equation}
where $a$ is related to dilatonic charge. In the following we
only consider the case where $\lambda > 0$. (Perhaps $\lambda =
0$ case can also be considered along the lines given in
\cite{bl}.) The mass $M_{sg}$ and the entropy $S_{sg}$ of the
corresponding charged black hole in the supergravity
approximation can be written as
\begin{equation}\label{sg}
M_{sg} = 2 b \left( \lambda \mu + \gamma \sqrt{Q^2 + \mu^2} 
\right) \; , \; \; \; 
S_{sg} = A \left(\sqrt{Q^2 + \mu^2} + \mu\right)^\gamma 
(2 \lambda b \mu)^\lambda 
\end{equation}
where $Q$ is the black hole charge, $A = \frac{4 \pi b}{d} \;
(\lambda b)^{- \lambda}$ and the constant $b$, which includes
brane tension and volume, can be obtained from expressions given
in \cite{kt'}. Note that upon defining $Q = \mu \; Sinh \; 2
\phi$, equations (\ref{sg}) become 
\begin{equation}\label{sgphi}
M_{sg} = 2 b \mu (\lambda + \gamma \; Cosh \; 2 \phi)
\; , \; \; \; 
S_{sg} = A (\lambda b)^\lambda 
(2 \mu)^{\lambda + \gamma} (Cosh \; \phi)^{2 \gamma} \; . 
\end{equation}
The above expressions describe $Dp-$branes for $(D, \gamma, a) =
(10, \frac{1}{2}, \frac{p - 3}{2})$, \\ 
$M-$branes for $(D, \gamma, a) = (11, \frac{1}{2}, 0)$, and
other branes for other values of $(D, \gamma, a)$: for example,
$(6, 1, 0)$ and $(5, \frac{3}{2}, 0)$. See \cite{kt'}.

In the extremal limit where $\mu = 0$, the mass and entropy are
given by $M_e = 2 b \gamma Q$ and $S_e = 0$ since $\lambda > 0$.
Thus, $Q$ can be taken to be the number $N$ of branes in the
stack, with $b$ containing the brane tension and volume
factors. In the near extremal limit where $\mu$ is small, the
brane dynamics can be obtained from the above solutions and can
be thought of as arising due to a gas of massless excitations.
Defining the energy $E$ of the gas on the branes to be $E =
M_{sg} - M_e \simeq 2 \lambda b \mu$, one obtains
\begin{equation}\label{sbeb}
S(E) = A N^\gamma E^\lambda \; . 
\end{equation}

We now extremise with respect to $N$ the total entropy $S_{tot}$
in (\ref{stot}), keeping the charge $q = N - \bar{N}$ and the
total mass $M$ fixed, and setting $E = \bar{E}$. Also, $C = 2 b
\gamma$. This then determines $N$ and $E$ to be given by 
\begin{equation}
2 E = M - 2 b \gamma (N + \bar{N}) = 4 \lambda b \; 
\frac{N^\gamma + \bar{N}^\gamma}
{N^{\gamma - 1} + \bar{N}^{\gamma - 1}}  
\end{equation}
and $\bar{N} = N - q$. For $\gamma = \frac{1}{2}$, the above
equations become 
\[
2 E = M - b (N + \bar{N}) = 4 \lambda b \sqrt{N \bar{N}}
\]
and can be solved for $N$, $\bar{N}$, and $E$ in terms of $M$
and $q$. \footnote{The above equations can be solved for $\gamma
= 1$ also. However, it turns out that the above analysis needs
to be generalised when $\gamma = \frac{K}{2}$, with $K$ an
integer $> 1$.} The solution can be parametrised as
\[
N = \frac{m}{2} e^{2 \theta} \; , \; \; \; 
\bar{N} = \frac{m}{2} e^{- 2 \theta} \; .
\]
Then, the required quantities can all be expressed in terms of $m$
and $\theta$. The result is:
\begin{eqnarray}
M & = & 2 b m (\lambda + \frac{1}{2} \; Cosh \; 2 \theta) \; , \; \; \; 
q = m \; Sinh \; 2 \theta \nonumber \\
S_{tot} & = & 2^{- \lambda} A (\lambda b)^\lambda 
(2 m)^{\lambda + \frac{1}{2}} \; Cosh \; \theta \; .
\end{eqnarray}
Comparing with the corresponding quantities in the supergravity
approximation after setting $\theta = \phi$ and $m = \mu$ so
that $M_{sg} = M$, it can be easily seen that $Q = q$ and
\begin{equation}
S_{tot}(M, q) = 2^{- \lambda} S_{sg}(M, q) \; . 
\end{equation}
This is essentially the description, given in \cite{dgk}, of
charged black hole in terms of branes and antibranes. 

{\bf 3.}
We now consider multicharged black holes. In the extremal and
near extremal limit, they can be described as intersecting
$p-$branes of String/M theory. Explicit solutions corresponding
to such multicharged black holes can be found in
\cite{ct,kt,ib}. We present here only the expressions for the
mass and the entropy of the multicharged black holes in the
supergravity approximation; they will suffice for our purposes
here. For complete details, see \cite{ct,kt,ib}. Denoting by $K$
the number of charges, the mass $M_{sg}$ and the entropy
$S_{sg}$ of the multicharged black holes are given by
\begin{eqnarray}
M_{sg} & = & 2 b \left( \lambda \mu + 
\sum_{i = 1}^K \gamma_i \sqrt{Q_i^2 + \mu^2} \right) 
\nonumber \\
S_{sg} & = & A \; \prod_{i = 1}^K 
\left(\sqrt{Q_i^2 + \mu^2} + \mu\right)^{\gamma_i} \;  
(2 \lambda b \mu)^\lambda \label{sgK}
\end{eqnarray}
where $Q_i$, $i = 1, 2, \cdots, K$ are the charges, the
constants $A$ and $b$ are given as before in (\ref{sg}) and
$\lambda$, assumed to be $> 0$ in the following, is now given by
\begin{equation}\label{lambdaK}
\lambda = \frac{d + 1}{d} - {\cal C} \; , \; \; \; 
{\cal C} = \sum_{i = 1}^K \gamma_i \; \; . 
\end{equation}
Note that upon defining $Q_i = \mu \; Sinh \; 2 \phi_i$,
equations (\ref{sgK}) become
\begin{eqnarray}
M_{sg} & = & 2 b \mu \left( \lambda + 
\sum_{i = 1}^K \gamma_i \; Cosh \; 2 \phi_i \right)
\nonumber \\
S_{sg} & = & A (\lambda b)^\lambda 
(2 \mu)^{\lambda + {\cal C}} \; \prod_{i = 1}^K 
(Cosh \; \phi_i)^{2 \gamma_i} \; .  \label{sgphiK}
\end{eqnarray}

In the extremal limit where $\mu = 0$, the mass and entropy are
given by $M_e = 2 b \sum_i \gamma_i Q_i$ and $S_e = 0$ since
$\lambda > 0$. Thus, $Q_i$ can be taken to be the number $\nu_i
\equiv N_i$ or $\bar{N}_i$ of branes or antibranes in the stack,
with $b$ containing the brane tension and volume factors. In the
near extremal limit where $\mu$ is small, the brane dynamics can
be obtained from the above solutions and can be thought of as
arising due to a gas of massless excitations. Defining the
energy $E$ of the gas on the branes to be $E = M_{sg} - M_e 
\simeq 2 \lambda b \mu$, one obtains
\begin{equation}\label{sbebK}
S(E) = A \; \left( \prod_{i = 1}^K \nu_i^{\gamma_i} \right) \;
E^\lambda \; . 
\end{equation}

We now describe the corresponding multicharged black holes as a
system consisting of stacks of intersecting branes and
antibranes, with a gas of massless excitation on each stack. In
supergravity approximation, a stack of intersecting branes
describes well the extremal and near extremal limits of
multicharged black holes which have zero entropy in the extremal
limit. The entropy $S(E)$ of the gas on the stack of branes is
then the same as that of the charged black holes in near
extremal limit, which is given by equation (\ref{sbebK}).

There is a subtlety now. Description of multicharged black holes
involves stacks of intersecting branes and antibranes, with
$N_i$ and $\bar{N}_i$, $i = 1, 2, \cdots, K$, being the number
of $i^{th}$ type of branes and antibranes. Let
\[
{\cal N}_I = (\nu_1, \nu_2, \cdots, \nu_K) 
\]
where $\nu_1 = N_1$, $\nu_i = N_i$ or $\bar{N}_i$ for $i = 2,
\cdots, K$, denote the numbers of constituent branes/antibranes
in a stack of intersecting brane configuration. We use ${\cal
N}_I$ to denote also the corresponding stack itself. The
subscript $I$, taken to be in the range $I = 1, 2, 3, \cdots,
2^{K - 1}$, denotes a particular realisation of $\nu_i$, $i = 2,
\cdots, K$. Let $\bar{\nu}_i = \bar{N}_i (N_i)$ when $\nu_i =
N_i (\bar{N}_i)$ and let
\[
\bar{{\cal N}}_I = (\bar{\nu}_1, \bar{\nu}_2, 
\cdots, \bar{\nu}_K) \; . 
\]
Thus, for example, if ${\cal N}_1 = (N_1, N_2, \cdots, N_K)$
then $\bar{{\cal N}}_1 = (\bar{N}_1, \bar{N}_2, \cdots,
\bar{N}_K)$.

Now, following DGK, multicharged black holes can be described as
a system consisting of two stacks, ${\cal N}_I$ and $\bar{\cal
N}_I$ for any single $I$, of intersecting branes and antibranes,
together with the gas of massless excitations. This pair of
stacks will have zero entropy, net charge $q_i = N_i -
\bar{N}_i$, $i = 1, 2, \cdots, K$, and mass $E_t$ due to
brane/antibrane tension given by
\[
E_t = 2 b \sum_{i = 1}^K \gamma_i (N_i + \bar{N}_i) \; . 
\]
Such a system, consisting of stacks of intersecting branes and
antibranes, will have gas(es) of massless excitations living on
them. In the single charged case DGK have argued, based on
physics involving tachyon condensation, that one copy of gas
lives on stack of branes and another on that of antibranes. In
the present case, where the system consists of a pair of stacks
${\cal N}_I$ and $\bar{\cal N}_I$ for a single $I$, the
corresponding tachyon physics is not clear. Certainly, as in
DGK, there should be one copy of gas on each of these
stacks. However, once the system is put in place, it can be
thought of (or ``taken apart'') as a pair of stacks with any
value of $I$, along with a copy of gas on it. Hence, we assume
that a copy of gas, with energy $E_I$ or $\bar{E}_I$ and entropy
$S_I$ or $\bar{S}_I$ corresponding to each stack ${\cal N}_I$ or
$\bar{\cal N}_I$, is present in the system for each value of
$I$. Thus, there are $2^K$ copies of gas in total which are
further assumed, following \cite{dgk}, not to interact with each
other.

The entropies $S_I$ and $\bar{S}_I$ are then obtained from the
supergravity description of near extremal dynamics and are given
by
\begin{equation}\label{si}
S_I = A \; \left( \prod_{i = 1}^K \nu_i^{\gamma_i} \right)_I \; 
E_I^\lambda \; , \; \; \; 
\bar{S}_I = A \; \left( \prod_{i = 1}^K \bar{\nu}_i^{\gamma_i} 
\right)_I \; \bar{E}_I^\lambda \;. 
\end{equation}
The subscript $I$ in the above expressions means that $\nu_i$'s
and $\bar{\nu}_i$'s are those corresponding to the stacks ${\cal
N}_I$ and $\bar{\cal N}_I$.

Thus, the total energy $M$ and the total entropy $S_{tot}$ of
the system of intersecting branes and antibranes and the $2^K$
copies of non interacting gas living on them are now given by
\begin{eqnarray}
M & = & 2 b \sum_{i = 1}^K \gamma_i (N_i + \bar{N}_i)
+ \sum_I (E_I + \bar{E}_I) \label{m'} \\
S_{tot} & = & \sum_I (S_I + \bar{S}_I)  \; . \label{stot'} 
\end{eqnarray}

In normal physical systems consisting of multiple components in
equilibrium, it is natural to assume that all the components are
at the same temperature. This is ensured by interactions between
the components, no matter how weak, and the principles of
statistical mechanics and ergodicity. It turns out that in the
present system, consisting of $2^K$ copies of gas, such an
assumption leads to results unconnected to charged black holes.
However, if we assume that the energies $E_I$ and $\bar{E}_I$ of
the gases are all equal to each other for all $I$, then the
resulting dynamics describes that of the charged black hole. We
will now assume this to be the case and proceed with the
analysis, and comment on it afterwards.

With this assumption, namely $E_I = \bar{E}_I = E$ for all $I$,
the total mass $M$ and the total entropy $S_{tot}$ of the system
now become
\begin{eqnarray}
M & = & 2 b \sum_{i = 1}^K \gamma_i (N_i + \bar{N}_i)
+ 2^K E \label{m} \nonumber \\
S_{tot} & = & A E^\lambda \; 
\prod_{i = 1}^K (N_i^{\gamma_i} + \bar{N}_i^{\gamma_i}) 
\label{mstot}
\end{eqnarray}
where we have used the following relation which follows easily :
\[
\sum_I \left( \prod_{i = 1}^K \nu_i^{\gamma_i} 
+ \prod_{i = 1}^K \bar{\nu}_i^{\gamma_i} \right)_I = 
\prod_{i = 1}^K (N_i^{\gamma_i} + \bar{N}_i^{\gamma_i}) \; .
\]

As in \cite{dgk}, in canonical formalism, such a system is
unstable, for sufficiently small values of $q_i$, towards
creating an infinite number of brane antibrane pairs; whereas,
it is stable in microcanaonical formalism for any value of
$q_i$. Hence, we work in microcanonical formalism where the
total energy $M$, and the charges $q_i \equiv N_i - \bar{N}_i$,
$i = 1, 2, \cdots, K$, of the system are kept fixed and the
equilibrium quantities are obtained by maximising the entropy
$S_{tot}$ of the system with respect to $N_i$. Hence, we 
maximise with respect to $N_i$ and $\bar{N}_i$ 
\[
S_{tot} + \sum_{i = 1}^K l_i (N_i - \bar{N}_i - q_i) 
\]
where $l_i$'s are Lagrange multipliers. After a straightforward
algebra, we get
\begin{eqnarray}
2^K E & = & M - 2 b \sum_{j = 1}^K \gamma_j (N_j + \bar{N}_j)
\nonumber \\
& = & 4 \lambda b \; 
\frac{N_i^{\gamma_i} + \bar{N}_i^{\gamma_i}} 
{N_i^{\gamma_i - 1} + \bar{N}_i^{\gamma_i - 1}}  
\label{maximise}
\end{eqnarray}
and $\bar{N}_i = N_i - q_i$ where $i = 1, 2, \cdots, K$. For
$\gamma_i = \frac{1}{2}$ for all $i$, the above equations become
\[ 
2^K E = M - b \sum_{j = 1}^K (N_j + \bar{N}_j) 
= 4 \lambda b \; \sqrt{N_i \bar{N}_i} 
\; , \; \; \; i = 1, 2, \cdots, K 
\] 
and can be solved for $N_i$, $\bar{N}_i$, and $E$ in terms of
$M$ and $q_i$.

Note that for all intersecting brane configurations in String
and M theories, the exponents $\gamma_i$ are indeed given by
$\frac{1}{2}$. Also, if the exponents are integer multiples of
$\frac{1}{2}$ then they can be obtained by String/M theory
intersecting branes by setting suitable number of charges to be
equal. Thus, for example, $\gamma = \frac{3}{2}$ in a single
charged case can be obtained from intersecting branes with $K =
3$ and setting $q_i = q$, $i = 1, 2, 3$. Indeed, this appears to
be the only way of obtaining such values of $\gamma$.
Interestingly, only such values of $\gamma$ appear in the known
cases \cite{kt',ct,kt,ib}. Hence, we set $\gamma_i =
\frac{1}{2}$, for all $i$, in the following.

The corresponding solutions can be parametrised as 
\[
N_i = \frac{m}{2} e^{2 \theta_i} \; , \; \; \; 
\bar{N}_i = \frac{m}{2} e^{- 2 \theta_i} \; .
\]
Then, the required quantities can all be expressed in terms of
$m$ and $\theta_i$. The result is:
\begin{eqnarray}
M & = & 2 b m (\lambda + \frac{1}{2} \; 
\sum_{i = 1}^K Cosh \; 2 \theta_i) 
\; , \; \; \; 
q_i = m \; Sinh \; 2 \theta_i \nonumber \\
S_{tot} & = & 2^{- \lambda K} A (\lambda b)^\lambda 
(2 m)^{\lambda + \frac{K}{2}} \; 
\prod_{i = 1}^K Cosh \; \theta_i \; .
\label{mstotphi}
\end{eqnarray}

Now compare with the corresponding quantities in the
supergravity approximation given in equations (\ref{sgphiK}),
with $\gamma_i = \frac{1}{2}$ and thus ${\cal C} = \sum_i
\gamma_i = \frac{K}{2}$. Setting $\theta_i = \phi_i$ and $m =
\mu$, we get $M_{sg} = M$, $Q_i = q_i$ and
\begin{equation}
S_{tot}(M, q_i) = 2^{- \lambda K} S_{sg}(M, q_i) 
\equiv X S_{sg}(M, q_i) \; . 
\end{equation}
Thus, the two entropies are equal upto a deficit factor $X$.

{\bf 4.}
To understand further the deficit factor $X$, we study how the
resulting total entropy $S_{tot}$ changes if one normalises the
brane tension, the gas energy, and entropy by constant factors,
as in \cite{k}.  For this, we consider the total energy and
entropy of the configuration to be given, under the same
assumptions as before, for example $E_I = \bar{E}_I = E$ for all
$I$, by
\begin{eqnarray}
M & = & 2 \alpha b \sum_{i = 1}^K \gamma_i (N_i + \bar{N}_i)
+ 2^K E \nonumber \\
S_{tot} & = & \sigma A (\epsilon E)^\lambda \; 
\prod_{i = 1}^K (N_i^{\gamma_i} + \bar{N}_i^{\gamma_i}) \; . 
\label{mstotscaling} 
\end{eqnarray}
The factor $\alpha$ normalises brane tensions,\footnote{ In
equation (\ref{mstotscaling}), $\alpha$ normalises only the
total brane energy which includes tensions, volumes, and number
of branes. However, it is natural to take $\alpha$ as
normalising brane tensions, and thereby the total brane energy.}
$\sigma$ the gas entropy, and $\epsilon$ the energy available to
each copy of the $2^K$ copies of the gas. Maximising the total
entropy $S_{tot}$ with respect to $N_i$, setting $\gamma_i =
\frac{1}{2}$, solving for $N_i$, etcetera as before, one obtains
the result
\begin{eqnarray}
M & = & 2 \alpha b m (\lambda + \frac{1}{2} \; 
\sum_{i = 1}^K Cosh \; 2 \theta_i) 
\; , \; \; \; 
q_i = m \; Sinh \; 2 \theta_i \nonumber \\
S_{tot} & = & \sigma \left(\frac{\epsilon}{2^K}\right)^\lambda 
A (\lambda \alpha b)^\lambda 
(2 m)^{\lambda + \frac{K}{2}} \; 
\prod_{i = 1}^K Cosh \; \theta_i \; . 
\label{mstotphiscaling} 
\end{eqnarray}
We define the supergravity charges $Q_i$ to be $Q_i = \alpha q_i
= \alpha (N_i - \bar{N}_i)$, or equivalently set $\mu = \alpha
m$. This should be so since $\alpha$ normalises brane tension
and, hence, the charges. Setting $\theta_i = \phi_i$ and
comparing with the corresponding supergravity quantities, we get
\begin{equation}
S_{tot}(M, Q_i) = X S_{sg}(M, Q_i) \; , \; \; \; 
X \equiv \sigma \alpha^{- \frac{K}{2}}
\left( \frac{\epsilon}{2^K} \right)^\lambda \; . \label{X}
\end{equation}
That this is the correct expression for the deficit factor, with
scalings included, can be checked explicitly for simple cases
like Schwarzschild black hole ($N_i - \bar{N}_i = 0$) or for a
single charged black hole ($K = 1$), by expressing the entropies
explicitly as a function of $M$ and $Q$. For example, for the
later case, one gets after some algebra
\begin{eqnarray}
S_{sg}(M, Q) & = & \; \; A \lambda^\lambda b^{- \gamma} 
\; \left(
\frac{\lambda - \gamma \sqrt{Z}}{\lambda - \gamma} 
\right)^\lambda \; 
\left( \frac{1 + \sqrt{Z}}{2} \right)^\gamma \; 
\left( \frac{M}{\lambda + \gamma} \right)^{\lambda + \gamma}
\nonumber \\
S_{tot}(M, q) & = & X \; A \lambda^\lambda 
b^{- \frac{1}{2}} \; \left(
\frac{\lambda - \gamma \sqrt{z}}{\lambda - \gamma} 
\right)^\lambda \; 
\left( \frac{1 + \sqrt{z}}{2} \right)^\gamma \; 
\left( \frac{M}{\lambda + \gamma} 
\right)^{\lambda + \gamma} \nonumber \\
X & \equiv & \sigma \alpha^{- \frac{1}{2}} 
\left( \frac{\epsilon}{2} \right)^\lambda
\end{eqnarray}
where 
$Z = 1 + \frac{4 b^2 Q^2 (\lambda^2 - \gamma^2)}{M^2}$, 
$z = 1 + \frac{4 \alpha^2 b^2 q^2 (\lambda^2 - \gamma^2)}{M^2}$, 
$K = 1$, and $q = N - \bar{N}$. $\gamma$ is arbitrary in the
expression for $S_{sg}$ and $ = \frac{1}{2}$ in that for
$S_{tot}$. With $Q = \alpha q$ and taking $\gamma =
\frac{1}{2}$, we get $Z = z$ and
\begin{equation}
S_{tot}(M, Q) =  X S_{sg}(M, Q) 
\end{equation}
which agrees with equation (\ref{X}). 

Clearly, the total entropy $S_{tot}$ of the intersecting brane
antibrane configurations will be exactly equal to the entropy
$S_{sg}$ of the corresponding multicharged black hole in
supergravity approximation if the deficit factor $X = \sigma
\alpha^{- \frac{K}{2}} \epsilon^\lambda 2^{- \lambda K} = 1$ for
any value of $\lambda$ and $K$.

Let $\sigma = \alpha = 1$ as in \cite{dgk}. Then the deficit
factor $X = 1$ for any value of $\lambda$ if $\epsilon =
2^K$. This is as if each copy of the gas carries $2^K$ times the
energy assumed to be available to it \cite{dgk}. If true, this
would violate energy conservation. 

However with $\sigma$ and $\alpha$ present and $\ne 1$, the
deficit factor $X = 1$ for any value of $K$ and $\lambda$, if we
set
\[ 
\epsilon = 2^K \; , \; \; \; 
\sigma = \frac{1}{2^K} \; , \; \; \; 
\alpha = \frac{1}{4} \; .  
\]
These values for $\epsilon, \sigma$, and $\alpha$ admit a simple
empirical interpretation \cite{k}. The value of $\epsilon$ means
that the gas energy is to be increased by a factor of $2^K$ and
the value of $\sigma$ means that the total gas entropy is to be
decreased by a factor of $2^K$. Empirically, they can be taken
together to mean simply that the available gas energy is not
shared equally by the $2^K$ copies of the gas but, instead, is
all taken by one single gas with its entropy given by the
average entropy of the $2^K$ copies of the gases. $\alpha =
\frac{1}{4}$ means that the brane tension is to be decreased by
a factor of $4$, which can perhaps be thought of as a net effect
of non trivial dynamics of intersecting branes and antibranes.

The precise nature of the single gas mentioned above is not
clear. In the case of the corresponding Schwarzschild black hole
\cite{k}, $N_i = \bar{N}_i$ and, hence, the $2^K$ copies of gas
are identical to each other\footnote{in the sense of having
identical entropy vs energy relation}, and also to the single
gas. Thus, this single gas can be taken to be one copy - or,
more generally, to be one linear combination - of the $2^K$
copies, which has all the available energy $= 2^K E$ in it.

In the case of charged black hole, the $2^K$ copies of the gas
are in general different from each other\footnote{in the sense
of having different entropy vs energy relation}, and also from
the single gas mentioned above. Then, this single gas can
perhaps be thought of as a gas which has all the available
energy $= 2^K E$ in it. Furthermore, it must perhaps be thought
of as sloshing back and forth as a whole between the $2^K$
stacks of branes/antibranes, spending equal time on each of the
stack and, thereby, having an entropy equal to the average
entropy of the $2^K$ copies of the gases. In this sense, this
single gas can be thought of as a certain superposition of all
the $2^K$ copies. However, the precise nature and dynamics of
the superposition is not clear to us.

If this interpretation is correct then the entropy of the brane
antibrane configuration is exactly equal to that of the
multicharged black hole in supergravity approximation. Another
attractive feature of this interpretation is the following. Note
that energy is now conserved since all the available energy is
taken by one single gas. Moreover, the assumption that energies,
not temperatures, of the different copies of the gas are
identical becomes superfluous. Such an assumption is unnatural
from a physical point of view, and is hard to realise
physically. But, in the present interpretation, it is rendered
superfluous.

It is conceivable that the above interpretation somehow captures
the essence of brane antibrane dynamics relevent for the
description of charged black holes. Then, understanding the
detailed properties of the single gas mentioned above will lead
to the description of the charged black hole. On the other hand,
the deficit factor $X$ may simply be due to the `binding energy'
of branes and antibranes, an interpretation advocated recently
in \cite{peet}. To understand these issues fully one needs a
rigorous study of brane antibrane dynamics at finite temperature
which, however, is likely to require the full arsenel of string
field theory techniques \cite{sftreview}.

{\bf 5.}
We described multicharged black holes using brane antibrane
configurations, generalising those in \cite{dgk}. The agreement
between the entropy of the intersecting brane antibrane
configurations and that of the corresponding multicharged black
hole in supergravity approximation is impressive. But these two
entropies differ by a deficit factor. We provided an empirical
interpretation of it.

We conclude by mentioning a few issues that can be studied
further. It will be interesting to understand the (near)
extremal dynamics of the multicharged black holes in terms of
the brane antibrane configurations, along with the gas of
massless excitations living on them. This is particularly so
since in the present interpretation, there is only one single
gas living on brane/antibranes (as opposed to two copies in
\cite{dgk} which play an important role in the (near) extremal
description).

More detailed description of multicharged black holes, such as
the emission and absorption cross section, requires a better
understanding of finite temperature brane antibrane
dynamics. String field theory techniques \cite{sftreview} are
essential towards a study of such issues. For some ideas in this
context, see \cite{dgk,peet} and references therein.

{\bf Note:} While this paper was being written, there appeared a
paper \cite{l} which has some overlap with the present one.



\end{document}